\documentclass{svjour3}                     
\smartqed  
\usepackage{graphicx}
\usepackage{color}
\usepackage{url} 

 \journalname{Scientometrics}
\begin{document}

\title{
Quantifying the evolution of a scientific topic: reaction of the
academic community to the Chornobyl disaster
%
}


\author{O.~Mryglod \and Yu.~Holovatch \and R.~Kenna \and B.~Berche }

\institute{O. Mryglod \at
              Institute for Condensed Matter Physics of the National Academy of Sciences of Ukraine,
              1 Svientsitskii Str., 79011 Lviv, Ukraine\\
              \email{olesya@icmp.lviv.ua}
           \and
            Yu. Holovatch \at
              Institute for Condensed Matter Physics of the National Academy of Sciences of Ukraine,
              1 Svientsitskii Str., 79011 Lviv, Ukraine
           \and
            R. Kenna \at
              Applied Mathematics Research Centre, Coventry University,
              Coventry, CV1 5FB, England
           \and
            B. Berche \at
               Universit\'e de Lorraine, Statistical Physics Group, IJL, UMR CNRS 7198, Campus de Nancy, B.P. 70239, 54506 Vand\oe uvre l\`es Nancy Cedex, France}

\date{Received:  / Accepted: date}

\maketitle

\begin{abstract}
We analyze the reaction of academic communities to a particular
urgent topic which abruptly arises as a scientific problem. To this
end, we have chosen the disaster that occurred in 1986 in Chornobyl
(Chernobyl), Ukraine, considered as one of the most devastating
nuclear power plant accidents in history. The academic response is
evaluated using scientific-publication data concerning the  disaster
using the Scopus database to present
the picture on an international scale and the bibliographic database
``Ukrainika naukova'' to consider it on a national level. We measured distributions of
papers in different scientific fields, their growth rates and
properties of co-authorship networks. {The elements of descriptive statistics and the tools of the complex network theory are used to highlight the interdisciplinary as well as international effects.} Our analysis allows to compare contributions of the international community to Chornobyl-related
research as well as integration of Ukraine in the international
research on this subject. Furthermore, the content analysis of
titles and abstracts of the publications allowed to detect the most
important terms used for description of Chornobyl-related problems.
\end{abstract}

\section*{Introduction}
Tracing the evolution of scientific topics is a subject area which belongs to the general problem of mapping the structure of scientific knowledge. It is important for distribution of funds, detection of `hot topics', strategic management  and other objectives.
Can we find the most promising areas for research?
Is it possible to explain  interest in particular topics or to predict one?
How should one monitor the evolution of topics which discuss  competitive
solutions of a problem? Due to a complexity of science, these and many other questions remain
open \cite{problems_1,problems_2,problems_3}. This case study is
an attempt to contribute to the exploration of these problems.

Similar to studying public opinion on a particular topic by
analysing news in mass media, a method to trace the usage of
scientific concepts is to analyse published works. The first
evidence for the emergence of a new topic is likely to be found in
the journal papers. Therefore, bibliometrical data bases can be used
to study the history of scientific topic evolution from its
appearance and until its extinction or merger with other topics
\cite{Zuccala11}. The results of such an analysis are presented in
this paper: our aim is to show how the scientific community reacts
to a particular problem, what are the collaboration profiles and the
interdisciplinary landscape, and whether one can detect a fading or
maintenance of interest to the problem. Unlike in \cite{Zuccala11},
where wide and continuously topical problem was chosen for analysis,
here we concentrate on the discussion of a particular event, which
occurred at a certain point in time. In this case we can precisely
define the starting point of the emergence of the new topic. The
fact that two of us are Ukrainian had some bearing on the choice of
this study too. {Of course, the reaction of society on particular events has been studied before as well. Several kinds of natural disasters including Chornobyl accident were considered as triggers to analyse the publication reaction in newspapers, Internet newsgroups, several kinds of scholar and technical bibliographic databases, etc., see \cite{Weessies07,Magnone12}. But different (more limited) set of tools were used in these studies. In particular, the analysis of reaction on Chornobyl accident in \cite{Weessies07} is more concentrated on the comparison of scholar and non-scholar data. Here we combine numerous technics to analyse academic response in more  details including interdisciplinary and international effects. Moreover, different databases are used here: Scopus database, which is considered as more interdisciplinary and more international comparing to Web of Science, is used for bibliometric data collection on a world-level. The data, that correspond to the local region of the accident are gathered from the Ukrainian abstract database called \emph{Ukrainika naukova} \cite{Ukrainika}. Such an attention to a particular area, which is geographically close to the event epicenter, is not our invention -- the same approach was used in \cite{Magnone12} to analyse the short-term effects in publishing activity after Japan's triple disaster.}

One of the most devastating nuclear accidents happened on April 26, 1986 at the Chor\-no\-byl nuclear power plant in Ukraine.
Chornobyl-related problems have been discussed during the last 29 years by society, including, of course, the
scientific community.
The evolution of the corresponding topic is studied here.
Our study is based on  \emph{Scopus} publication data, supplemented by  bibliographic information retrieved from the Ukrainika naukova.

The description of our datasets along with the results of initial statistical analysis is given in the section~\ref{data}.
The international collaboration profile is examined in section~\ref{collaboration}, while the interdisciplinary landscape and subtopics diversity are discussed in
sections~\ref{interdisciplinary} and \ref{terms}, correspondingly.
Some of our results were previously announced in \cite{Mryglod12}.

\section{Datasets and their evolution}
\label{data}

All the publications from \emph{Scopus}, which contain any spelling of ``Chor\-no\-byl'' (\emph{chornobyl OR chornobyl' OR chernobyl OR chernobyl'}) in their titles, abstracts or keywords were chosen for analysis: above 9\,500 records in total\footnote{The
data collected during January--February 2015 are used in this
research.}.
Only a very small number  --- just 19 papers --- were published before the accident.
This indicates that the vast majority of publications are related to this particular event and we consider to the publication period starting from 1986.
The searching request for the \emph{Ukrainika naukova} database was performed in a similar way:
different spellings ``Chornobyl'' in English, Ukrainian and Russian
 were used to find the relevant publications.
Although the latter is far from being complete, it is the best
source of information about the Ukrainian- and Russian-language
publications in local journals and books, especially for social
sciences and humanities. At the end, 1\,918 items,
starting from 1997, were found. Another peculiarity of
\emph{Ukrainika naukova} is that single record can refer to not only
single journal papers, but also to collection of publications, book
or conference proceedings.

\begin{figure}[ht]
\centerline{\includegraphics[width=0.5\textwidth]{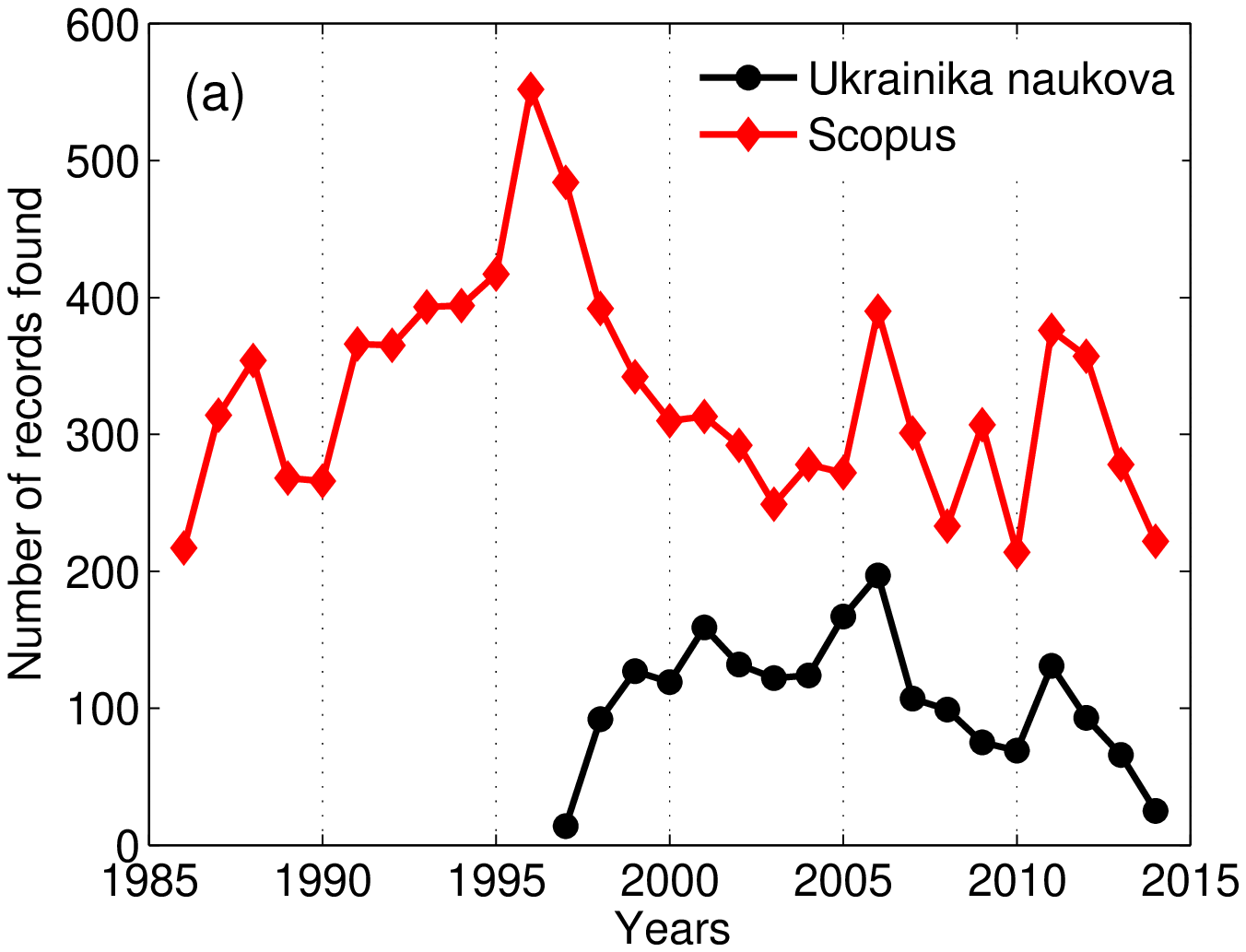}
\hfill
\includegraphics[width=0.5\textwidth]{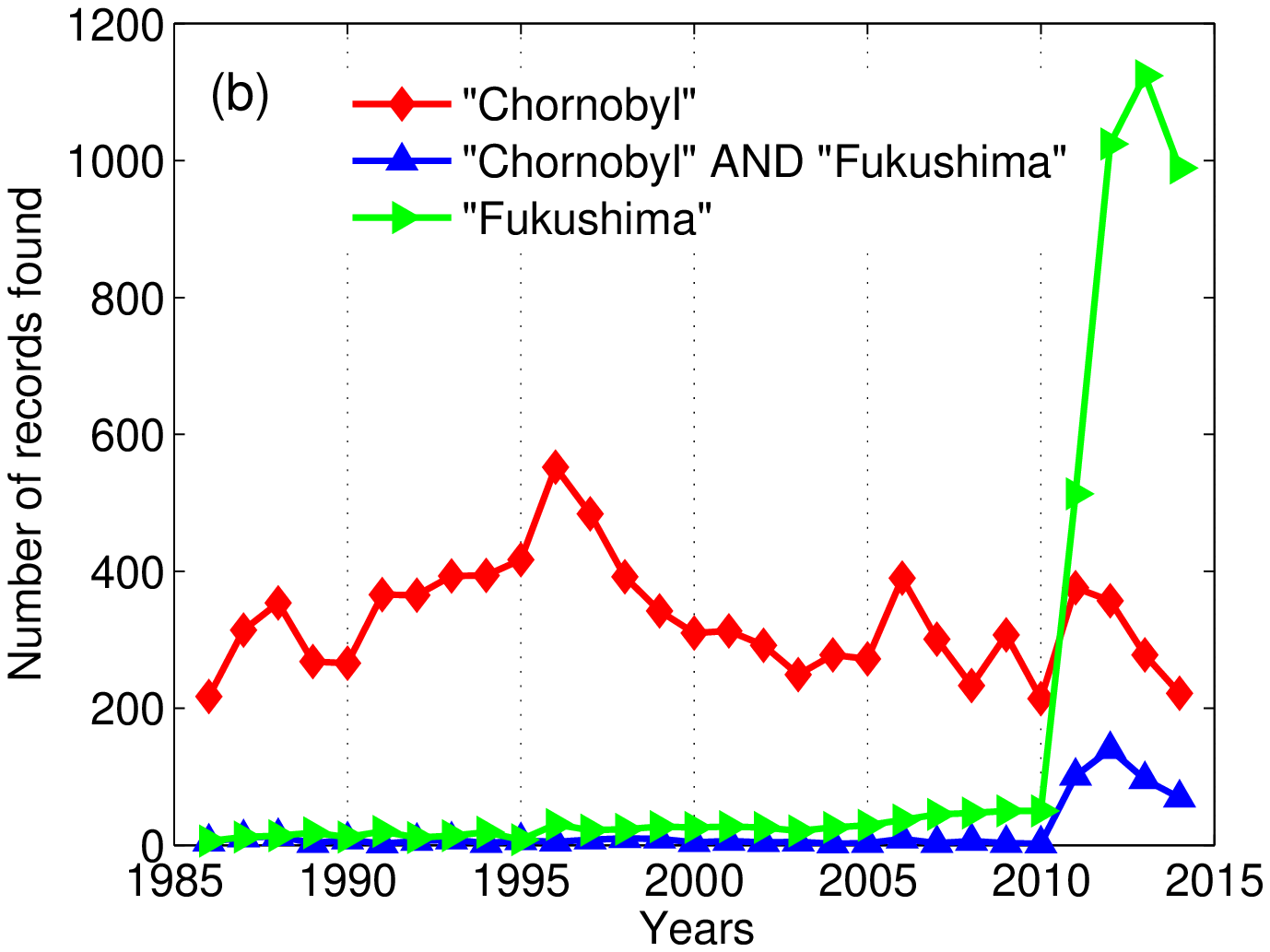}}
\caption{Number of search results (a) from \emph{Scopus} and \emph{Ukrainika naukova} by `Chornobyl' request; (b) from \emph{Scopus} by `Chornobyl', `Chornobyl AND Fukushima' and `Fukushima' requests.}
\label{fig0}
\end{figure}

In Fig.~\ref{fig0} (a) the annual total number of Chornobyl-related
records from \emph{Scopus} and \emph{Ukrainika naukova} are shown.
The peaks which occurred due to an `anniversary effect' can
be instantly distinguished: in 1996 (10 years after the accident)
for \emph{Scopus} data, in 2006 and in 2011 (20th and 25th
anniversaries, correspondingly) for both datasets {-- the same effect was observed earlier in \cite{Weessies07}}.
There is a smaller peak in the \emph{Scopus} data in 2009 and we assume that this is partly attributable to a number of publications in the Annals of the New York Academy of Sciences \cite{translation} which
appeared as a  collection of  translated chapters of
Russian book about Chornobyl, which was originally published in
2007.
{The corresponding publication peaks are also observed for the separate disciplines, as it will be seen further from Figs.~\ref{fig_kovb_01} and \ref{fig_kovb_02} (see also \cite{Weessies07}). This gives us a possibility to compare our results for Chornobyl accident case with the results, obtained for the other topic, which is not characterized by special time limits. In \cite{Zuccala11} the dynamics of poverty research is analysed and no particular peaks similar to those observed in Figs.~\ref{fig_kovb_01}, \ref{fig_kovb_02}  can be observed in figure 1 from this paper. This allows to assume that such distinct patterns of publication activity are rather not characteristic to the gradually emerging topic, but more typical for particular event-related problems.}

One  notices also for \emph{Scopus} data, that the decay after 2011 is
less steep comparing to the ones which followed  previous peaks
(Fig.~\ref{fig0} (a) and (b)). This can be explained by the
 nuclear catastrophe in 2011 in Japan (Fukushima);  the appearance of the competitive topic presumably
provokes the number of comparative studies and the reactivation of
interest to Chornobyl.
On the other hand, the new problem naturally
attracts more attention causing the fading of major interest in the
older one. In Fig.~\ref{fig0} (b) one can clearly see the increased
number of papers relevant to both Chornobyl and Fukushima in 2012,
but the total number of Fukushima-related papers is much larger.
Comparing the curves related to both accidents, it is easy to see
that the same patterns appear at the very beginning: number of
publications increases during the first three years after the
trigger event and decreases afterwards.

In section~\ref{collaboration} the analysis of international
collaboration within Chornobyl problem is considered, while the
object of research: interdisciplinary landscape and thematic
spectrum -- is studied in the last two sections
\ref{interdisciplinary} and \ref{terms}.

\section{International collaboration profile}
\label{collaboration}

The level of publication activity, analysed above, is an indicator
of general interest in the Chornobyl topic. Let us look on it now in
more details, specifying how this interest evolved in different
countries. To this end, we use  affiliation data of
Chornobyl-related papers to investigate  international
collaboration patterns. Unfortunately, not all publication records
contain the necessary information. The affiliation data are absent
for the majority of documents found in the \emph{Ukrainika naukova}.
However, \emph{Scopus} records are provided by such data and thus,
we continue our analysis  based only on this dataset.

\begin{figure}[ht]
\centerline{\includegraphics[width=8cm]{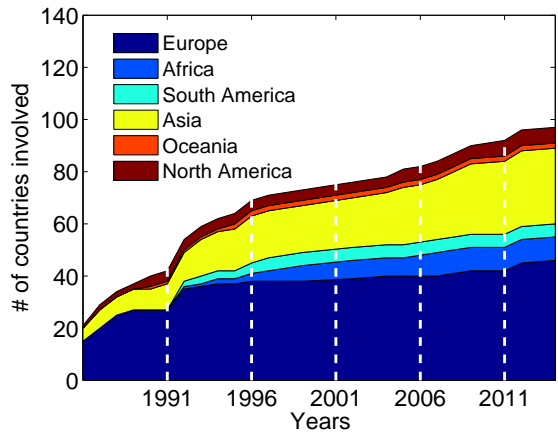}}
\caption{Cumulative number of countries from different parts of the world, involved in Chornobyl related research.}
\label{fig_part_world}
\end{figure}

In Fig.~\ref{fig_part_world} the cumulative numbers of countries from different parts of the world
which contributed scientifically to the Chornobyl problem are shown
for each year. Numerous European and Asian countries started to
contribute into Chornobyl topic from the very beginning, while some
countries joined after 1991 and then after 1995 {(the small ``jumps'' of cumulative curves can be observed in figure for these years)}.

\begin{figure}[ht]
\centerline{\includegraphics[width=0.98\textwidth]{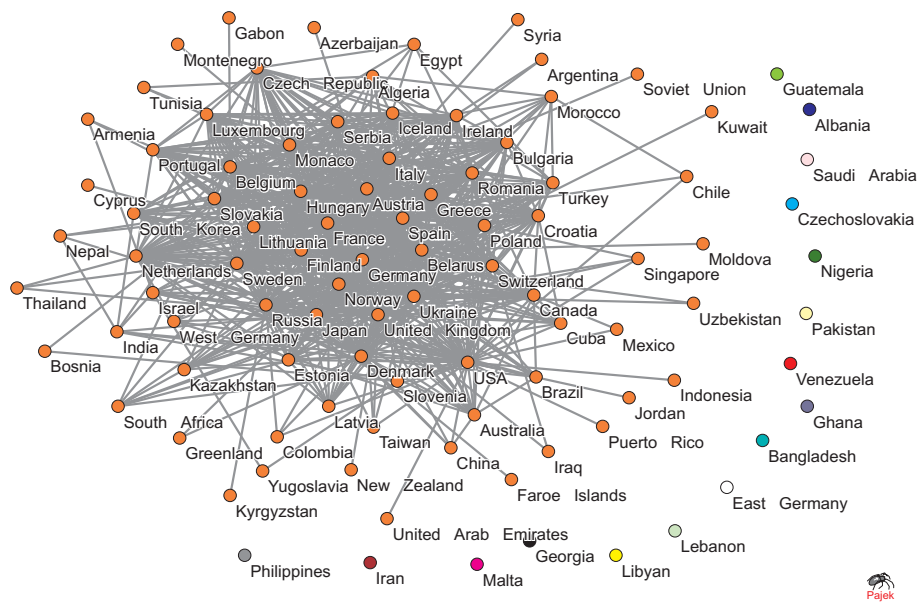}}
\caption{Collaboration network of countries for Chornobyl-related
research based on \emph{Scopus} data (publication period:
1986--2015). The nodes, which belong to the same component are shown
in a same color online. For the sake of better visualization, the
network is shown here without taking into account the weights of the links as
well as to the contribution share (i.e., number of papers) of
countries. This and further networks were generated using Pajek
network visualization software~\cite{Pajek}. } \label{fig_netw_all}
\end{figure}

Affiliation data are used to build the co-authorship network on
level of countries and, thus, to analyse the international
collaboration patterns. Nodes in this network represent different
countries, and these are deemed linked if a paper is co-authored by
authors from the two corresponding countries. For the weighted
variant of the network, the strength of each link is proportional to
the number of common papers. The network is visualized in
Fig.~\ref{fig_netw_all} {applying Pajek network visualization software~\cite{Pajek}, which is widely used for visualising and analysing complex networks, see, e.g., \cite{Leydesdorff14}}. It consists of 97 nodes, 80 of which
(almost 83\%) belong to a single connected (giant) component. Within
this component it takes 2 steps on average to connect any pair of
countries by collaboration links; the longest distance between two
countries consists of 4 steps. The average node degree, i.e., number
of adjacent links, is 15, which means that each country on average
collaborates with 15 others. Among the hubs with more than 40 links
two categories of countries can be identified -- the developed ones,
such as the USA, Germany, France, Austria, etc. and the countries
close to the event epicenter such as Ukraine, Russia and Belarus. On
the other hand, 17 isolated nodes correspond to the countries with
no collaboration within Chornobyl topic. This includes a number of
Asian and African countries, as well as countries that do not exist
anymore as political ebtities, such as Czechoslovakia. The value of
the network clustering coefficient, which shows the general tendency
to create cliques\footnote{{Fully connected graph (each node is connected with any other node) is characterized by clustering coefficient equals to 1, while value 0 characterize tree-like structure.}} (the fully connected network fragments), is close
to 0.78. This value, together with short average shortest path
between the pair of nodes and the fat-tailed degree distribution,
tells us that the constructed network is not random -- it is
compact, but strongly correlated.

\begin{figure}[ht]
\centerline{\includegraphics[width=0.7\textwidth]{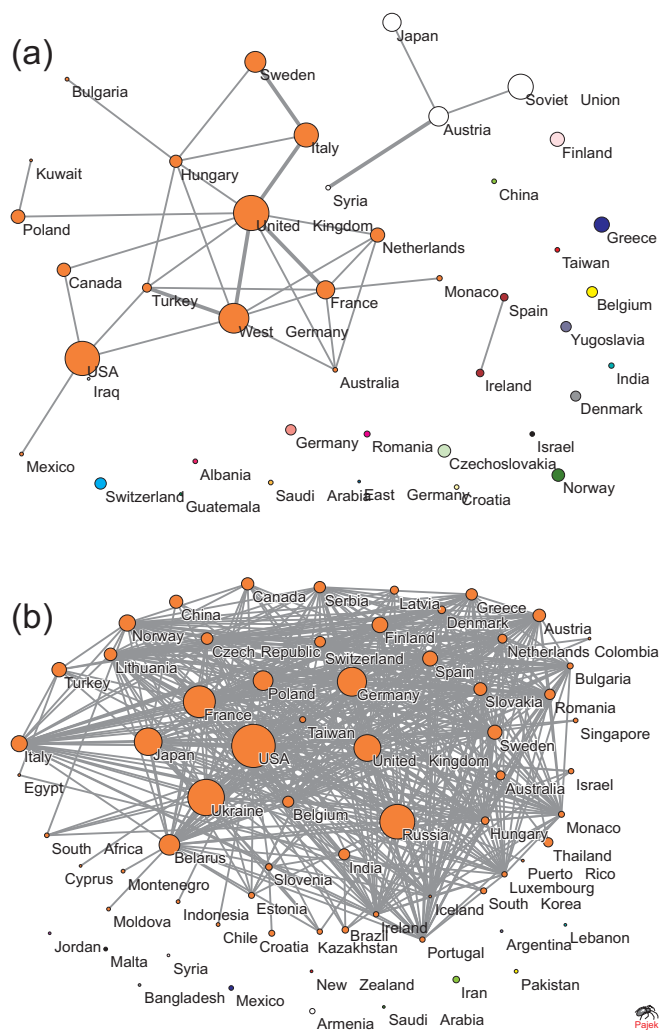}}
\caption{Weighted collaboration network of countries for
Chornobyl-related research based on \emph{Scopus} data, publication
period: (a) 1986--1991, (b) 2010--2015. The size of a node is
proportional to the total number of papers for the corresponding
country.} \label{fig_netw_first_and_last}
\end{figure}

To take into account the historical circumstances and to see how the
network dynamically changes in time, it is reasonable to study it
within particular time windows. To give an example, the
networks, generated for the first and the last five years after
Chornobyl disaster, are shown in Fig.~\ref{fig_netw_first_and_last}.
The network in
Fig.~\ref{fig_netw_first_and_last}~(a) is less
connected comparing to that Fig.~\ref{fig_netw_first_and_last}~(b):
the clustering coefficient is $\approx 0.46$ for the first period
and $\approx 0.82$ for the second. Besides, the size of the largest
component also increases, containing  only 38\% of nodes in the
first instance and 82\% in the second.
The reason for such densification is not completely clear.
It may be because science is becoming more collaborative
(due to, e.g., an improved communication environment thanks to the Internet and numerous international exchanging projects).
One can also suggest that the global consequences of the Chornobyl accident became more
obvious even for remote countries and, therefore, more studied
internationally after 20 years. To try to shed light on this and
other patterns of publication activity associated with the  Chornobyl problem, we
study the interdisciplinary profile as well as thematic spectra in
the following sections.

\section{Interdisciplinary landscape}
\label{interdisciplinary}

While the changing of the total number of Chornobyl-related
publications shows the general interest to the topic, it is hard
to say if the same dominant disciplines are `responsible' for all
publications peaks, or maybe interdisciplinary landscape is changing
over time. In order to determine this, further we analyse the
contribution of each discipline.

Since the problem of subject classification of scientific papers is
far from being trivial, the widely adopted way to assign the
disciplinary label is based on some metadata such as authors or
editorial keywords, or thematic indices such as PACS or UDC numbers
\cite{PACS,UDC}. Besides other drawbacks, this procedure usually
deals not with individual documents, but with their collections:
journals, books, etc.
On the one hand, according to the
\emph{Scopus} classification scheme, each journal title is assigned
to one or more subject categories. This gives us the opportunity to
divide the set of papers among 27 disciplines. On the other
hand, each record found in \emph{Ukrainika naukova} is labeled by
one or more special hierarchical subject indices, according to the
inner classifications scheme with 34 disciplines defined. Using the
implemented classification algorithms and having all these cautions in
mind, we perform the analysis of the interdisciplinary landscape of
Chornobyl-related publications.

\begin{figure}[ht]
\centerline{\includegraphics[width=1.1\textwidth]{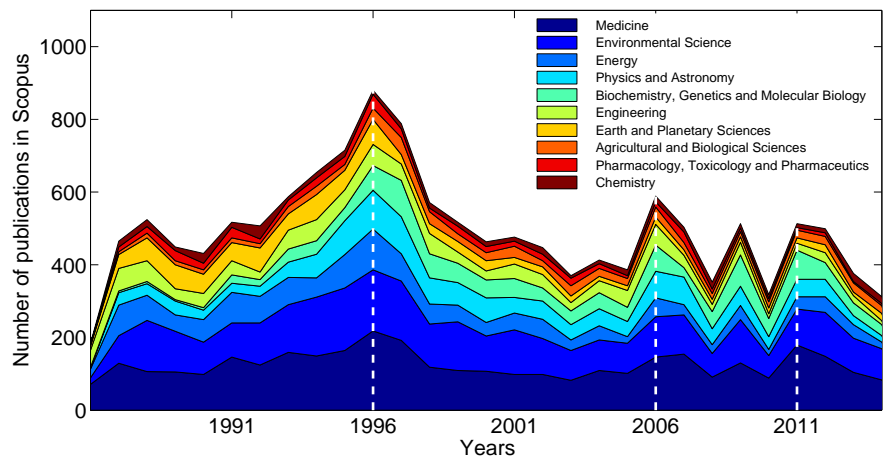}}
\caption{TOP10 of dominated subject areas by publication level based on the \emph{Scopus} data.}
\label{fig_kovb_01}
\end{figure}

\begin{figure}[ht]
\centerline{\includegraphics[width=1.1\textwidth]{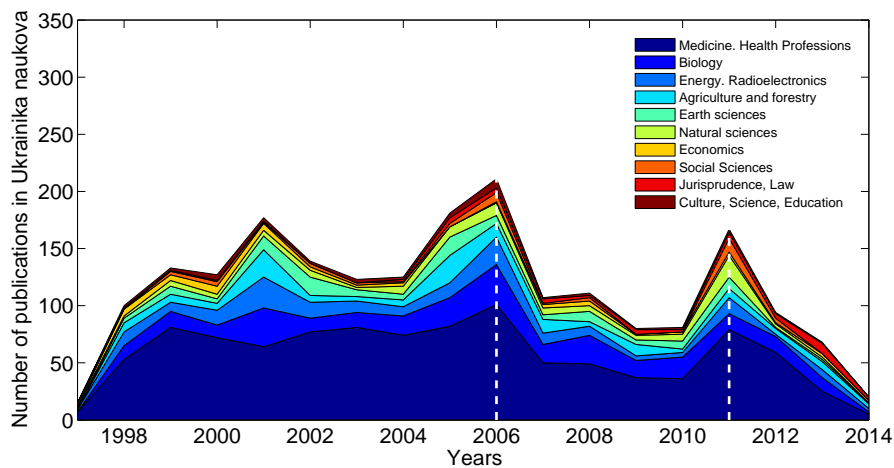}}
\caption{TOP10 of dominated subject areas by publication level based on the \emph{Ukrainika naukova} data.}
\label{fig_kovb_02}
\end{figure}

The current TOP5 list of disciplines by the total number of
Chornobyl-related papers in \emph{Scopus} contains: Medicine (3\,635
papers); Environmental Science (3\,156); Energy (1\,470); Physics
and Astronomy (1\,437); Biochemistry, Genetics and Molecular Biology
(1\,198). The annual numbers of publications, assigned to these and
the next five dominant disciplines in \emph{Scopus} are presented in
Fig.~\ref{fig_kovb_01}. Considering these data as representative
sample which reflects the international scientific reaction, one can
conclude that, naturally, the topics of highest interest relate to
the most problematic consequences of Chornobyl disaster: impact on
human health and the environment. A similar conclusion can be drawn
from Fig.~\ref{fig_kovb_02}, where the annual numbers of documents
in \emph{Ukrainika naukova} are shown. Six dominant disciplines here
also deal with medical and environmental problems. However, the next
four subject categories can be attributed to the other dimensions of
human life: Chornobyl-related problems in Economics, Social
Sciences, Law and Culture are highly discussed on the local scale.
The disproportion of \emph{Scopus} database coverage of natural and
social sciences on one hand and peculiarities of Ukrainian database
on the other hand can play a role here. But obviously, territories
which are close to the accident epicenter experience a higher
diversity of perturbations in different spheres of human life.
%
%
\begin{figure}[!h]
\centerline{\includegraphics[width=0.5\textwidth]{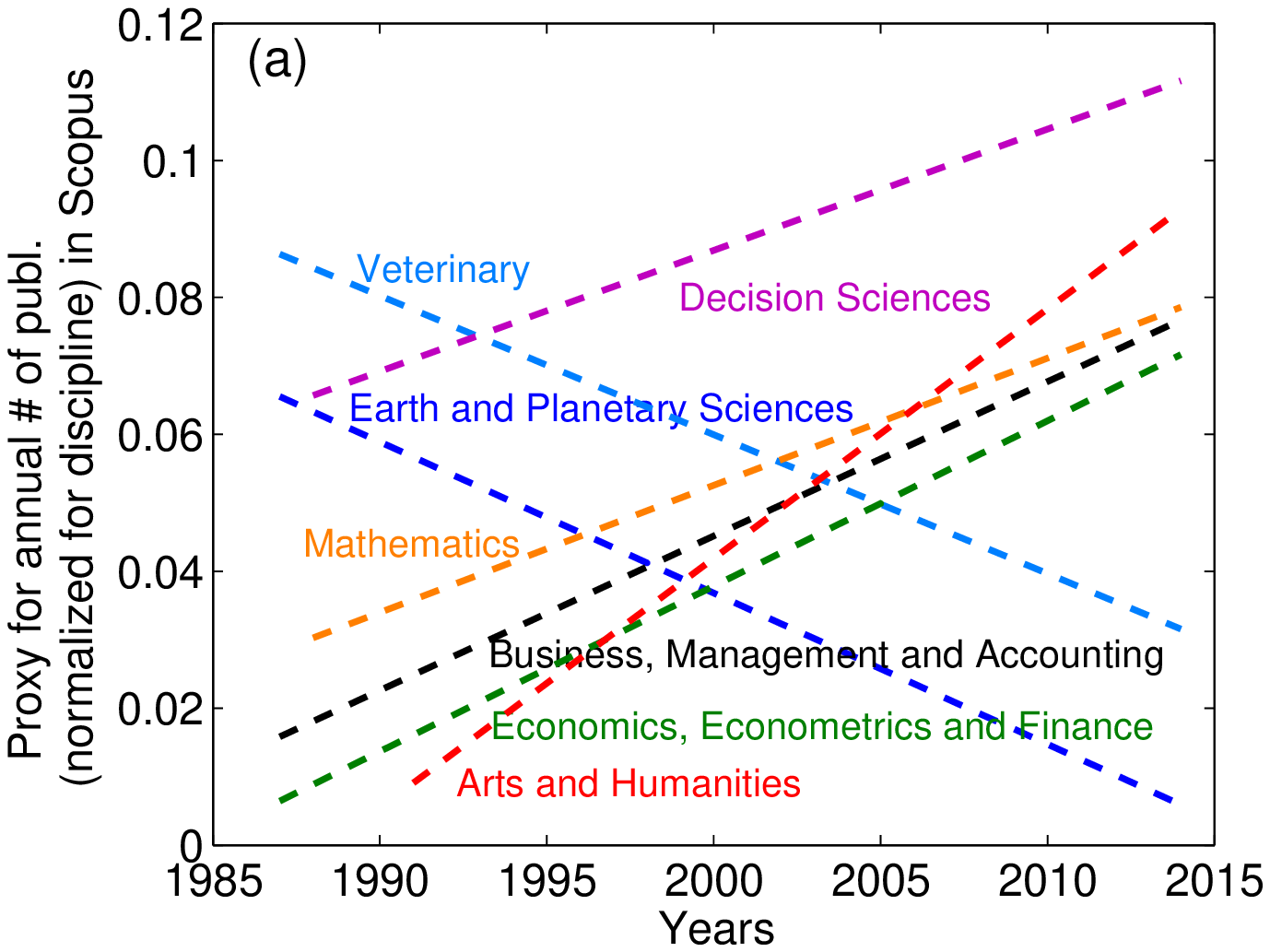}
\hfill
\includegraphics[width=0.5\textwidth]{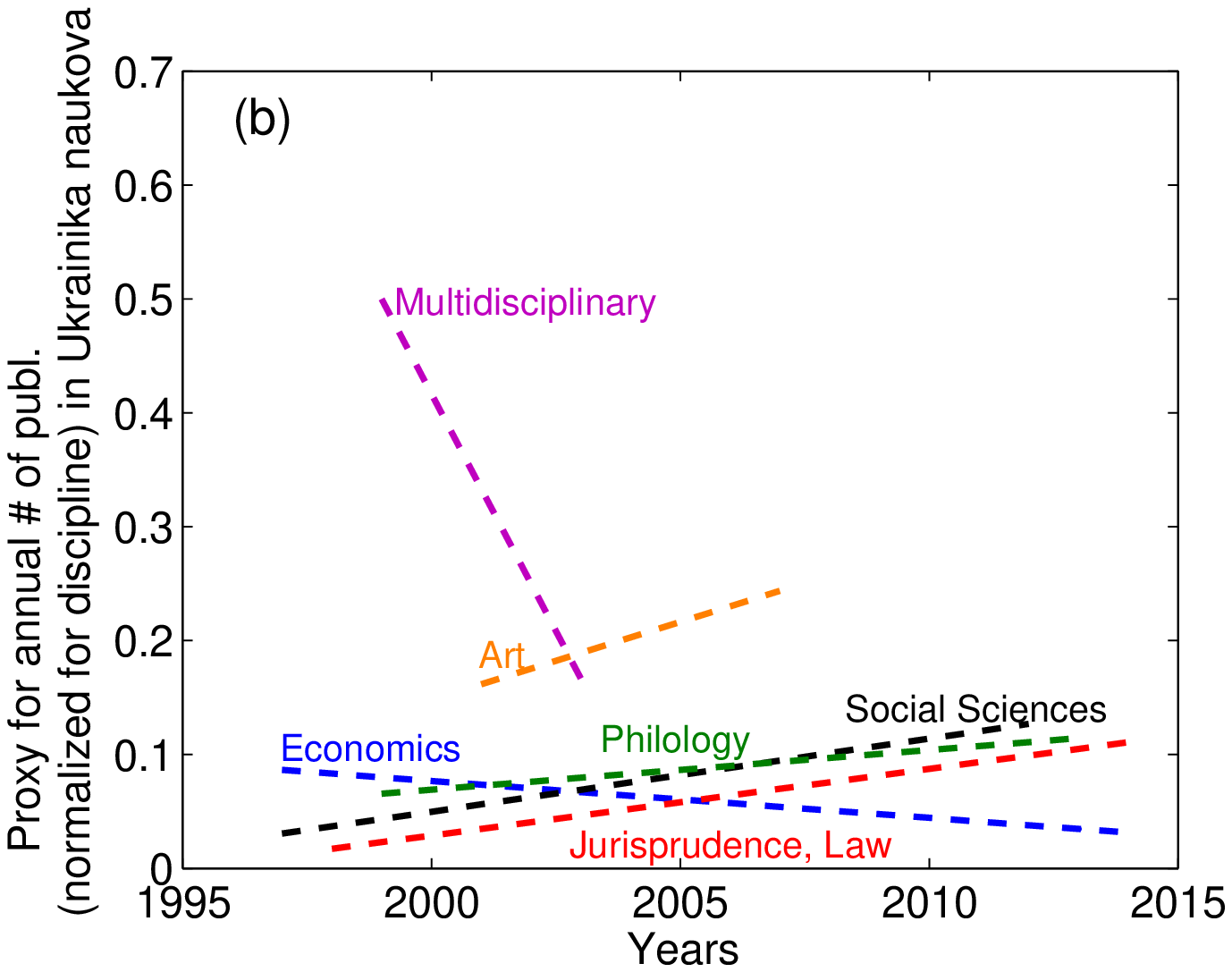}}
\caption{The `tendencies' built for a number of disciplines based on (a) \emph{Scopus} data, (b) \emph{Ukrainika naukova} data.}
\label{fig_tendencies}
\end{figure}

Figs.~\ref{fig_kovb_01} and \ref{fig_kovb_02} also show that the total amount of publications for some disciplines, such as Medicine, remains essentially stable during
the years, while the others, for example, Biochemistry, Genetics and
Molecular Biology, start to develop later. In Fig.~\ref{fig0}
one can notice that  the annual total numbers of documents
currently fluctuate on a certain level, indicating the sufficiently
constant interest to the topic on average. Yet, it is interesting to
investigate the tendencies within each particular discipline: each
line in Fig.~\ref{fig_tendencies} is a linear fit for the annual
number of published documents within a particular discipline,
normalized on the total number of documents for this
discipline\footnote{The years with no publications are not taken
into account here.}. The slopes of the lines indicate the `tendency'
within each particular discipline: negative one means that Chornobyl
problem becomes less discussed and vice versa; small absolute value
of  slope indicates a stable interest in the topic. Eventually,
the `immediate' consequences of Chornobyl disaster become less
discussed, while researchers switch to the study of `remote'
effects. To give an example, the initial interest to Chornobyl
problems in Veterinary fades out during 25 years, while such
disciplines as Arts or Economics become more and more relevant, see
Fig.~\ref{fig_tendencies} (a). Another good example, which can be
seen in Fig.~\ref{fig_tendencies} (b), is Law --
supposedly, due to a legal issues connected with a huge number of
Ukrainians moved from Chornobyl to the others regions and with the
families of accident liquidators, the legal problems are highly
discussed on a national scale.

Only 25\% of disciplines are represented in Fig.~\ref{fig_tendencies}.
These have the most significant absolute values of slope. The rest, including Medicine,
Environmental Science, Physics and Astronomy,
demonstrate rather a more stable level of interest in Chornobyl problems.
More detailed analysis such as an attempt to understand the topical
spectra of Chornobyl-related publications within particular
disciplines based on the content-analysis of titles or abstracts, is
discussed in section~\ref{terms}.

\section{Thematic spectrum}
\label{terms}

While the interdisciplinarity of Chornobyl-related research,
discussed in the previous section, gives a general picture of the
thematic structure, there are different approaches to investigate it
at a finer level \cite{Schneider_thesis,Eck_thesis,Eck10,Anick99}.
One possibility is to perform  content analysis dealing with the
full texts of publications, separate paragraphs, titles, abstracts
and/or keywords \cite{Zuccala11,Pollack15}. Here we connect with
scientometric problems called ``mapping of science'',  broadly
described in the literature starting from the second half of last
century \cite{mapping_old} and continuing with the most recent
research \cite{mapping_new}. Such maps show the thematic structure
of documents (scientific publications in our context) by grouping
the \emph{terms} in accordance to their similarity: the idea
 is that related or similar words appear
close to each other on a map.
The mapping problem is complex since  \emph{similarity} can be defined in different ways:
based on the number of co-occurrences in the documents or its
fragments, co-citations, or other data, Jaccard coefficient, Cosine
similarity, Euclidean distance, or others proximity measures can be
calculated (e.g., see \cite{Schneider_thesis}). On the other hand,
the notion of \emph{term} is not formalized so far. Usually,
\emph{terms} mean content words, which transmit the main ideas of
the document unlike the functional ones, which are used rather as a
`linguistic environment' to carry out the content words
\cite{Schneider_thesis,Pazienza05,Kageura96}. It is not a trivial
task to distinguish the former from the latter, therefore, various
approaches can be considered, but often the idea of distributional
differences between the content and functional words is explored
\cite{Schneider_thesis,Eck_thesis,Eck10,Anick99,Kageura96}. For
example, the distributions of occurrences or co-occurrences of words
in documents (or in the selected groups of documents) can be
compared. The higher values of the so-called termhood, which is
usually the numerical representation of such a difference, allows
one to label words as terms. Further we discuss this idea more in
details.

\begin{figure}[ht]
\centerline{\includegraphics[width=0.98\textwidth]{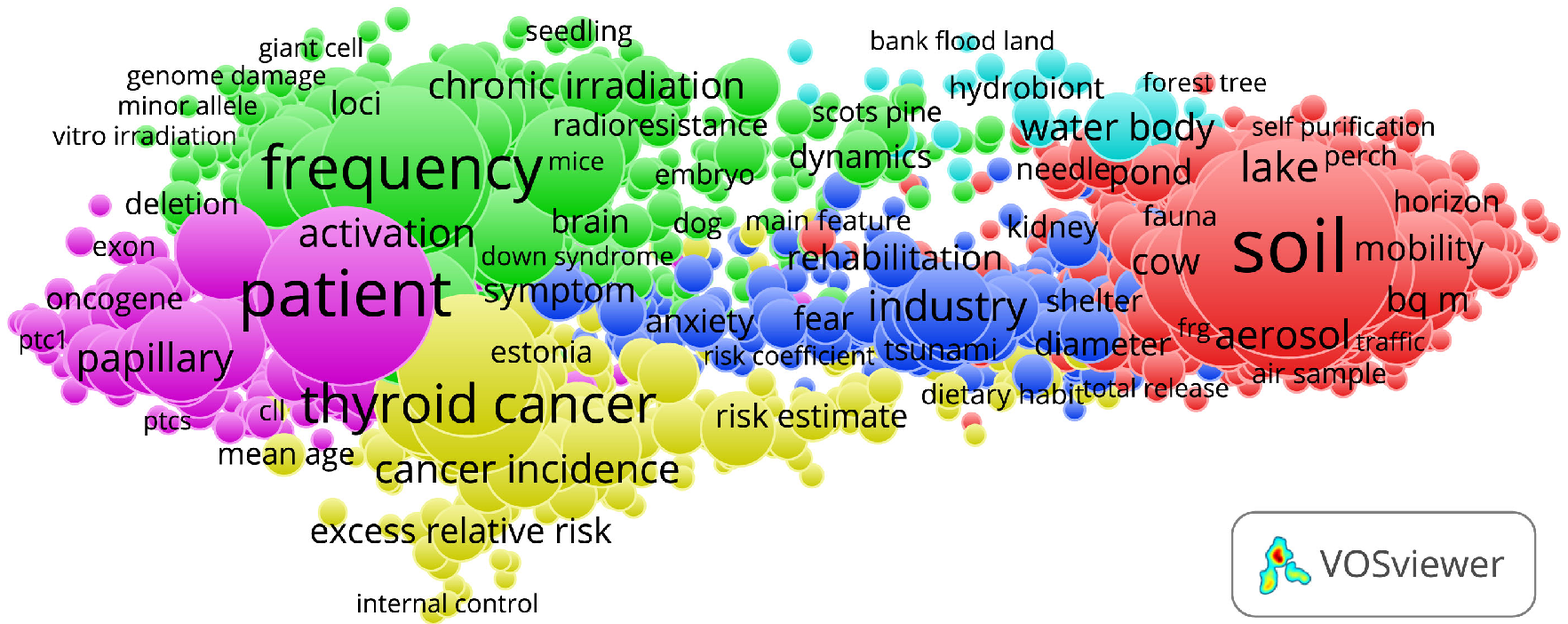}}
\caption{Term map of Chornobyl-related publications in TOP5 disciplines: Medicine; Environmental Science; Energy; Physics and Astronomy; and Biochemistry, Genetics and Molecular Biology.}
\label{fig_VOS}
\end{figure}

The terms map for Chornobyl-related papers within TOP5 disciplines
(see section~\ref{interdisciplinary}) found in \emph{Scopus}
database is shown in Fig.~\ref{fig_VOS}. This map is automatically
generated by VOSviewer software \cite{VOS,Eck_text} and it is based
on the analysis of co-occurrences of terms in the titles and
abstracts (each pair of title and abstract is treated as a single
document). There, the 60\% most relevant terms\footnote{Only words
and phrases, which occurred at least 10 times (default value) in
different documents were considered as the candidates for terms.}
for Chornobyl publications in TOP5 disciplines are mapped.
%
%
The clustering algorithm, implemented into the VOSviewer program \cite{Eck_community}, allows us to see several groups of terms (different colors online): four large and two smaller ones. While these clusters do not correspond one-to-one to the list of disciplines, one can clearly distinguish two main areas of Chornobyl-related problems which were already mentioned above: impact on human health (three big clusters on the left-hand side, purple, yellow and green online) and the environmental consequences (large cluster on the right-hand side, red online). Moreover, the dominant subtopics can also be seen from the first glance at the map: cancer threats, genetic effects, different ways of contamination spreading.

Taking one step back, here we describe the key points of
the term selection procedure we have chosen to
analyze the \emph{Scopus} set of Chornobyl-related publications.

\begin{itemize}
\item[i.] We analyse separately titles and abstracts, i.e.,  each title is
treated as one \emph{document} in the first case, while in the
second case \emph{document} contains only publication abstract.
\emph{Scopus} publication data for five dominant disciplines
(Medicine -- 3610 papers; Environmental Science -- 3141; Energy --
1476; Physics and Astronomy -- 1441; and Biochemistry, Genetics and
Molecular Biology -- 1192) and two additional ones (Arts and
Humanities -- 59 papers; Social Sciences -- 310) were used.
\item[ii.] Special linguistic software called tagger is used in the second step
to label each word by the part of speech and to
define its root form. As in \cite{Zuccala11,Eck10}, the freely
available program TreeTagger \cite{Schmid94,Schmid95,TreeTagger} is
used.
\item[iii.] Representing the `conceptual units' of the text, terms usually are expressed
in a form of noun phrases (single- or multiple-words)
\cite{Schneider_thesis,Eck_thesis,Eck10}. Similarly to \cite{Eck10},
as candidates for terms we select all single nouns and sequences of
words which satisfy the general rule:
    \verb``*adjective *noun''. This general form means, that a phrase can
start with any number of adjectives (optional) and ends with any
number of nouns, e.g. `genetic effect', `low-dose radiation
exposure', `invisible radioactive radiation', etc. The listed noun
phrases (both single- and multiple-words) -- our candidates for
terms -- are called \emph{semantic units} \cite{Eck_thesis,Eck10}.
In general, we get 80\,094 semantic units for abstracts dataset and
15\,020 for titles dataset.
\item[iv.]
We use the binary way to count the number of occurrences $k$ of
semantic units in the documents (each unit is
counted once per document). For multi-word semantic units we count
the number of occurrences $k$ not only for the entire semantic unit,
but also for its constituent parts. In this way we are dealing with
the so-called nested terms, which are terms that are part of other
longer terms \cite{Eck_thesis}. Example: for the semantic unit
`chernobyl nuclear power plant accident' we count the occurrences of
`chernobyl nuclear power plant accident'; `nuclear power plant accident';
`power plant accident'; `plant accident' and 'accident'. \\[1ex]
    Obtained in this way frequency-rank distributions for
    semantic units in abstracts and titles are shown in Figs.~\ref{fig_Zipf}~(a), (b), correspondingly.
    One can notice the power-law nature of these distributions, which is typical for the distribution of words in natural texts \cite{Zipf_in_text1,Zipf_in_text2}. Moreover, the power-law exponents in Fig.~\ref{fig_Zipf} are relatively close to 1 (especially for the abstracts dataset), demonstrating the well-known Zipf's law \cite{Zipf_def}. This is not an obvious result, since Zipf's law was formulated for the set of any words in the text, while here we operate only with noun phrases.
%
\begin{figure}[ht]
\centerline{\includegraphics[width=0.5\textwidth]{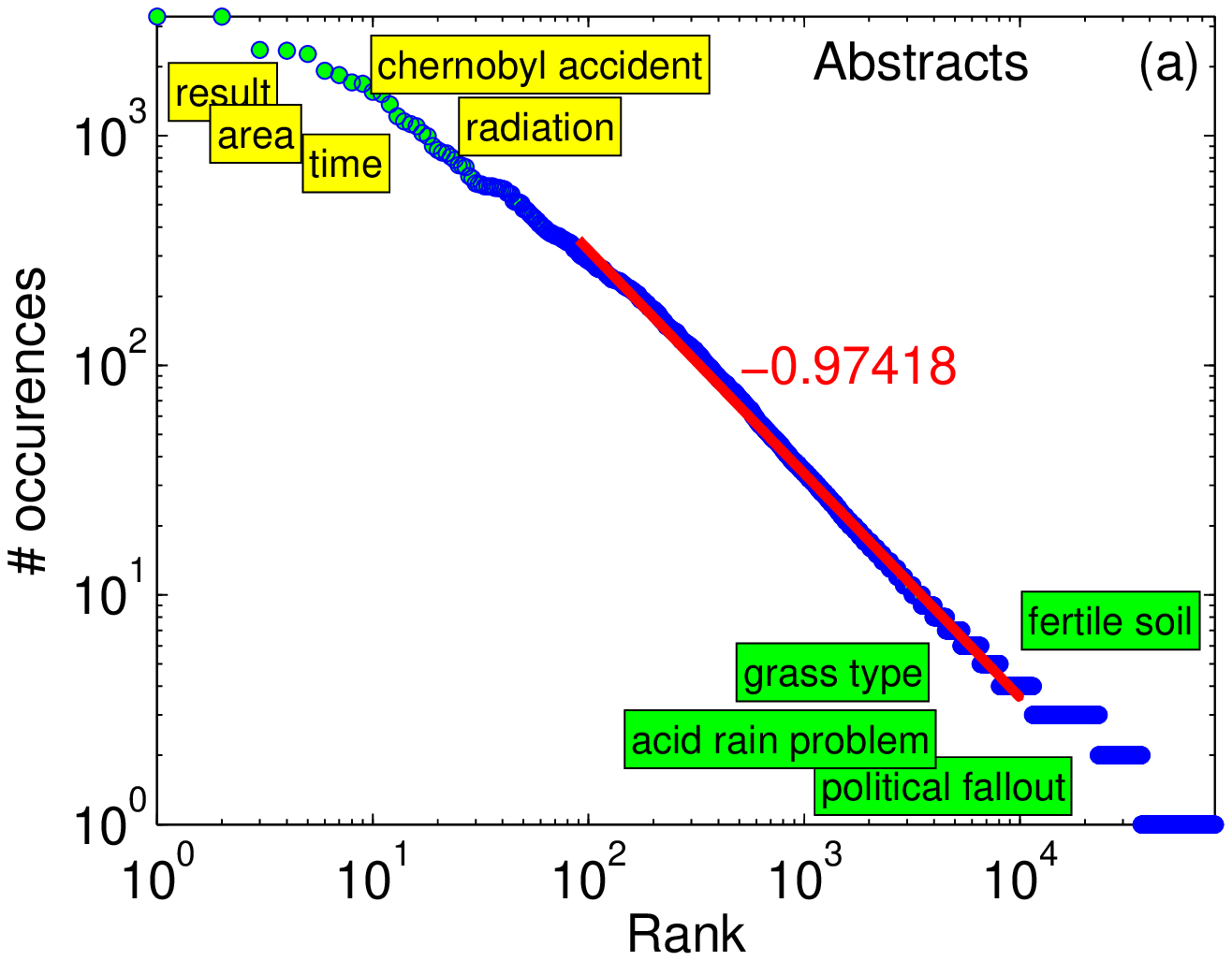}
\hfill
\includegraphics[width=0.5\textwidth]{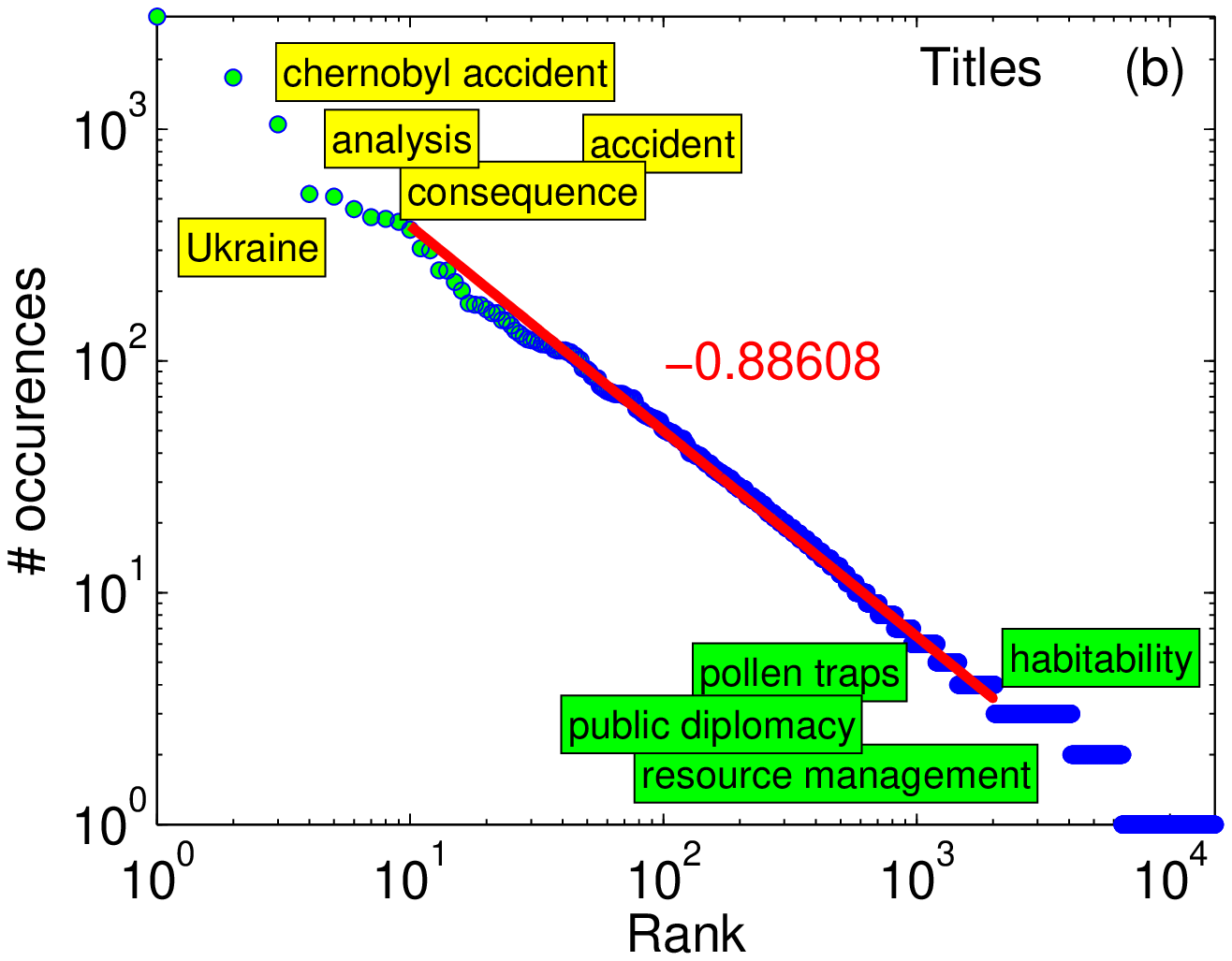}}
\caption{Frequency-rank distributions for semantic units in: (a)
abstracts and (b) titles of Chornobyl-related publications within
seven chosen disciplines based on the \emph{Scopus} data.}
\label{fig_Zipf}
\end{figure}
%
%
\item[v.] The terms should be defined and screened out from the final list of semantic units. Contrary to the intuitive understanding, the specificity of the term is not directly correlated with its frequency (this can be  also seen further in Fig.~\ref{fig_termhood}). It is known that the most frequent words, which are at the beginning of Zipf's plot, are usually the most general ones such as `Chornobyl accident' or `radiation' in our datasets (see Fig.~\ref{fig_Zipf}). The words with the lowest number of occurrences, which can be found in the tail of the distributions in Fig.~\ref{fig_Zipf}, are naturally very specific, but no statistical conclusions can be drawn. The most probable candidates for terms correspond to the middle part of the frequency-rank distribution \cite{Schneider_thesis}. Thus, we already apply the lower threshold for the number of occurrences $k$: semantic units, which occurred in four or less documents, were excluded from the analysis on this step. The choice of the critical value $k_{\mathrm{c}}=4$ was made because of an empirical observation that the number of semantic units with the number of occurrences $k$ above criticality decays slowly for $k_\mathrm{c}\leq 4$ and rapidly for $k_\mathrm{c}>4$.

Unlike a functional word, which can be distributed over topics or
disciplines equally randomly, a content word or term is supposed to
be \emph{specific}. The variant of \emph{termhood}, defined in
\cite{Eck_thesis,Eck10}, is used here to measure the specificity of
each semantic unit, i.e., how much it is biased towards a particular
discipline/disciplines. The termhood $t_j$ of a
semantic unit $s_j$ is defined as \cite{Eck_thesis,Eck10} a difference between
two probability distributions $P(d_i)$ and $P(d_i|s_j)$: $P(d_i)$
shows the probability of \emph{any} semantic unit to occur within
$i^{\mathrm{th}}$ discipline $d_i$ ($i=1\dots 7$), while
$P(d_i|s_j)$ provides the probability of $j^{\mathrm{th}}$ semantic
unit $s_j$ to occur within the $i^{\mathrm{th}}$ discipline $d_i$.
We illustrate both distributions in Fig.~\ref{fig_P(di)} where
$P(d_i)$ obtained for the abstract dataset is accompanied by the
distributions for two selected semantic units with high and low
termhood. {As the frequency of occurrences of term ``population'' within each discipline is close to the generally expected ones (the shape of $P(d_i|\mathrm{``population''})$ is close the shape of $P(d_i)$), we conclude that this particular term is not specific for any of disciplines considered. But the situation is different for the term ``cold war'': as one can see, it occurred much more frequently in humanity papers, that it is expected. Therefore, we consider it as specific for ``Arts and Humanities'' for our database. In our opinion, these findings correspond to the intuitive speculations quite well. The general shape of $P(d_i)$ is similar} for the titles dataset too.
%
\begin{figure}[ht]
\centerline{\includegraphics[width=0.8\textwidth]{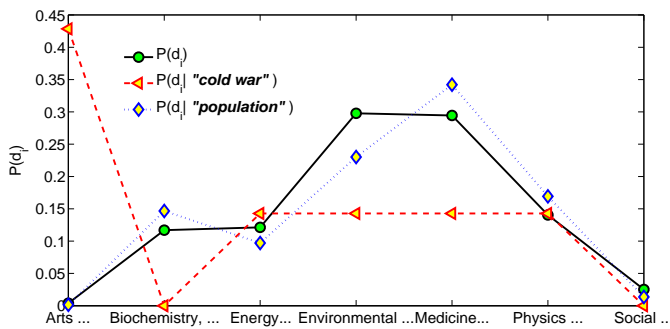}}
\caption{Probability distributions of semantic units occurrences
within different disciplines $P(d_i)$. General distribution is
accompanied by the distributions for two selected
semantic units: ``cold war'' (high termhood) and
``population'' (low termhood).} \label{fig_P(di)}
\end{figure}
    There are different methods of numerical comparison of two distributions but we continue to use the one suggested in \cite{Eck_thesis,Eck10}. Thus, the termhood of semantic unit $s_j$ is calculated in a following way:
    \[
    t_j=\sum_{i=1}^7{\log{p_i}}\,,
    \]
    where
    \[
     p_i=\frac{P(d_i|s_j)/P(d_i)}{\sum_{i'=1}^7{P(d_{i'}|s_j)/P(d_{i'})}}\,,
    \]
    and $0\log{0}=0$.

The higher the value of a termhood, the more specific (more relevant
to particular disciplines) term is. As one can see in
Fig.~\ref{fig_termhood}, the compromise between the frequency $k$
and specificity of the term $t$ should be found, but a solution is
not obvious. Therefore, any resulting list of terms will be
provisional to some extent. But since text mining procedures usually
requires expert evaluation on different stages, so we consider this
as normal. We end here with the list of terms, which satisfies
condition: $t_j>t_{\mathrm{c}}$, where $t_{\mathrm{c}}$ is the 50th
percentile (P50); and belongs to the TOP50 of the terms list, sorted
by values $t'_j \cdot k'_j$ (product of $t_j$ and $k_j$,
renormalized to fit them into [0..1] interval). Such a way of
ordering was chosen to give additional preference to specificity of
terms comparing to their frequency.
\begin{figure}[ht]
\centerline{\includegraphics[width=0.8\textwidth]{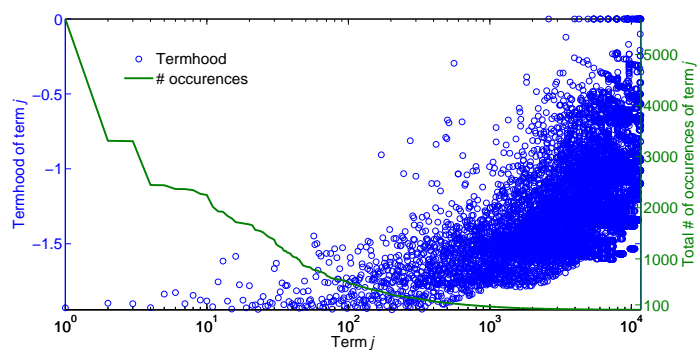}}
\caption{Termhood $t$ vs. total number of occurrences
$k>k_\mathrm{c}$ for semantic units in the abstracts dataset.}
\label{fig_termhood}
\end{figure}
\end{itemize}

The final TOP20 list of terms (see Tables~\ref{tab1} and
\ref{tab2}), obtained from abstracts and titles, contains mainly
those, which are more specific for Biochemistry, Genetics and
Molecular Biology, which occur at first time at the beginning of
90-ies. The selected environmental terms correspond to earlier years (1987--1991), while humanity dominant terms --
to later period (2002--2006). This brings us again to the idea of
non-synchronous development of Chornobyl topic within different
disciplines, as it was already shown above (e.g., see
Fig.~\ref{fig_tendencies}). {It is important to note, that much more interesting results could be obtained using cumulative data in order to perform the dynamical content analysis. Certainly, it would be useful to consider the occurrences frequency changes to see how particular terms become more or less topical in time. This can be done using larger database to operate with statistically sufficient numbers.}
\begin{table}[ht]
\caption{TOP20 list of terms, obtained from the abstracts
dataset.}
\begin{center}
\begin{tabular}{|l|l|l||l|l|l|}
\hline
Terms&Specific for Discipline:&\parbox[t]{1.5cm}{Year of the 1st occurrence} \\
\hline
\hline
carcinoma&\parbox[t]{5.5cm}{Biochemistry, Genetics and Molec. Biology}&1992\\
thyroid carcinoma&\parbox[t]{5.5cm}{Biochemistry, Genetics and Molec. Biology}&1992\\
tumor&\parbox[t]{5.5cm}{Biochemistry, Genetics and Molec. Biology}&1994\\
gene&\parbox[t]{5.5cm}{Biochemistry, Genetics and Molec. Biology}&1987\\
rearrangement&\parbox[t]{5.5cm}{Biochemistry, Genetics and Molec. Biology}&1993\\
papillary thyroid carcinoma&\parbox[t]{5.5cm}{Biochemistry, Genetics and Molec. Biology}&1995\\
ptc$^{\dag}$&\parbox[t]{5.5cm}{Biochemistry, Genetics and Molec. Biology}&1995\\
papillary carcinoma&\parbox[t]{5.5cm}{Biochemistry, Genetics and Molec. Biology}&1992\\
science&Arts and Humanities&2002\\
carcinogenesis&\parbox[t]{5.5cm}{Biochemistry, Genetics and Molec. Biology}&1994\\
malignancy&\parbox[t]{5.5cm}{Biochemistry, Genetics and Molec. Biology}&1992\\
activity ratio&Environmental Science&1987\\
threat&Arts and Humanities&2006\\
metastasis&\parbox[t]{5.5cm}{Biochemistry, Genetics and Molec. Biology}&1992\\
surgery&\parbox[t]{5.5cm}{Biochemistry, Genetics and Molec. Biology}&1994\\
policy&Social Sciences&1989\\
cleanup worker&\parbox[t]{5.5cm}{Biochemistry, Genetics and Molec. Biology}&1997\\
high frequency&\parbox[t]{5.5cm}{Biochemistry, Genetics and Molec. Biology}&1990\\
nuclear disaster&Social Sciences&1990\\
discharge&Environmental Science&1988\\ \hline
\end{tabular} \label{tab1}

\small{$^{\dag}$ ptc is the abbreviation of `papillary thyroid carcinoma'}
\end{center}
\end{table}

\begin{table}[ht]
\caption{TOP20 list of terms, obtained from the titles dataset.}
\begin{center}
\begin{tabular}{|l|l|l||l|l|l|}
\hline
Terms&Specific for Discipline:&\parbox[t]{1.5cm}{Year of the 1st occurrence} \\
\hline
\hline
carcinoma&\parbox[t]{5.5cm}{Biochemistry, Genetics and Molec. Biology}&1993\\
thyroid carcinoma&\parbox[t]{5.5cm}{Biochemistry, Genetics and Molec. Biology}&1993\\
patient&\parbox[t]{5.5cm}{Biochemistry, Genetics and Molec. Biology}&1991\\
rearrangement&\parbox[t]{5.5cm}{Biochemistry, Genetics and Molec. Biology}&1991\\
sediment&Environmental Science&1987\\
papillary thyroid carcinoma&\parbox[t]{5.5cm}{Biochemistry, Genetics and Molec. Biology}&1995\\
transport&Environmental Science&1987\\
mutation&\parbox[t]{5.5cm}{Biochemistry, Genetics and Molec. Biology}&1989\\
tumor&\parbox[t]{5.5cm}{Biochemistry, Genetics and Molec. Biology}&1994\\
cleanup worker&\parbox[t]{5.5cm}{Biochemistry, Genetics and Molec. Biology}&1993\\
cleanup&Medicine&1992\\
unit&Energy&1982\\
history&Arts and Humanities&2009\\
pond&Environmental Science&1987\\
policy&Social Sciences&1988\\
prevalence&\parbox[t]{5.5cm}{Biochemistry, Genetics and Molec. Biology}&1995\\
black sea&Environmental Science&1987\\
thyroid disease&\parbox[t]{5.5cm}{Biochemistry, Genetics and Molec. Biology}&1991\\
radiation protection&Arts and Humanities&2006\\
forest ecosystem&Environmental Science&1991\\ \hline
\end{tabular} \label{tab2}
\end{center}
\end{table}

\section{Summary}
The aim of this case study is to contribute to the complex problem
of analysing the scientific trends. To study the reactions of
scientific community to the particular topic -- Chornobyl disaster
-- the set of relevant publications was analysed. To this end a
number of different tools were used: bibliometrical analysis,
descriptive statistics methods, complex networks theory and the
elements of content analysis.

Summarizing the results, we can state that Chornobyl problem
permanently attracts the interest of researchers. While some
disciplines demonstrate a decreasing publication activity within
Chornobyl topic, the experts in other fields become more
interested. This can be explained by the gradual changing of
thematic spectrum: the environmental and medical consequences are
obvious from the very beginning, while remote genetic effects could
be seen later. The topics of current interest vary not only in with
respect to the disciplines, but also to the geographic remoteness
from the epicenter of the accident. While the international
scientific community all over the world is interested in medical,
environmental or physical problems related to Chornobyl, in the
neighbouring countries different social, cultural and other humanity
y aspects became also evident. Such countries, along with the most
developed ones, contribute the most. The collaboration within
Chornobyl-related problems becomes more and more tight: geography of
the problem expands. The content-analysis of publication abstracts
and titles allows one to refine analysis of thematic structure of
Chornobyl problem. Among numerous subtopics, two big sets of
problems remain the most important: those which are connected with
the human health and the environmental ones.

To conclude, the study of the evolution of competitive scientific topics would be an interesting challenge for the future. Having the results of analysis
of retrospective data, it is possible to try to predict the future
publication activity patterns. The results of numerous case studies
can be used in detecting the typical patterns, universal for
different topics.

\section*{Acknowledgements}
This work was supported by the 7th FP, IRSES projects No. 295302
``Statistical physics in diverse realizations'',  No. 612707
``Dynamics of and in Complex Systems'', No. 612669 ``Structure and
Evolution of Complex Systems with Applications in Physics and Life
Sciences'' and  by the COST Action TD1210 ``Analyzing the dynamics
of information and knowledge landscapes''. OM would like to thank to
Nees Jan van Eck for a useful discussion and explaining some key
features of the VOSviewer program.

\end{document}